\documentclass[preprint]{aastex}
\def\sss{\scriptscriptstyle}
\def\^#1{^{\sss #1}}
\def\_#1{_{\sss #1}}
\def\beq{\begin{equation}}
\def\eeqno#1{\label{#1}\end{equation}}
\def\ten#1#2{^{\sss#1}_{\sss#2}}
\def\TP{\textsf{P}}

\def\az{a\_{0}}

\def\l0{\ell\_{0}}

\def\rar{\rightarrow}

\def\l{\lambda}

\def\rp{\rho_p}
\def\f{\phi}
\def\fN{\phi\^N}
\def\gfN{\grad\fN}
\def\fs{\f^*}
\def\gfs{\grad\fs}
\def\vgN{\vg\^N}

\def\k{\kappa}
\def\z{\zeta}
\def\eN{\eta\_N}

\def\r{\rho}
\def\hr{\hat\rho}
\def\m{\mu}
\def\n{\nu}
\def\Up{\Upsilon}
\def\C{\Gamma}
\def\F{\mathcal{F}}
\def\L{\mathcal{L}}

\def\O{\mathcal{O}}
\def\P{\mathcal{P}}
\def\Q{\mathcal{Q}}
\def\V{\mathcal{V}}

\def\N{\mathcal{N}}
\def\M{\mathcal{M}}
\def\O{\mathcal{O}}

\def\D{\Delta}

\def\d{\delta}
\def\drt{d^3r}
\def\a{\alpha}
\def\b{\beta}
\def\c{\gamma}
\def\d{\delta}
\def\eps{\epsilon}
\def\vr{{\bf r}}

\def\vF{{\bf F}}
\def\vT{{\bf T}}

\def\vv{{\bf v}}

\def\vg{{\bf g}}

\def\vA{{\bf A}}
\def\vf{{\bf f}}
\def\vF{{\bf F}}

\def\S{\Sigma}
\def\vds{{\bf d\sigma}}

\def\grad{\vec\nabla}
\def\div{\vec \nabla\cdot}
\def\gf{\grad\phi}
\def\fpg{4\pi G}
\def\RM{R\_M}
\def\curl{\nabla\times}
\def\hnz{\hat\n_0}
\def\gmn{g\_{\m\n}}
\def\Gmn{g\^{\mu \nu}}

\def\gij{g\_{ij}}
\def\Gij{g\^{ij}}

\def\hgmn{\hat g\_{\m\n}}
\def\hgh{\hat g^{1/2}}
\def\gh{g^{1/2}}
\def\LP{L\_P}
\def\gft{\grad\tilde\f}

\def\vgt{\tilde\vg}
\def\ft{\tilde\f}
\def\hC{\hat\C}

\begin{document}

\title{Quasi-linear formulation of MOND}
\author{Mordehai Milgrom}
\affil{ The Weizmann Institute Center for Astrophysics}

\begin{abstract}
A new formulation of MOND as a modified-potential theory of gravity
is propounded. In effect, the theory dictates that the MOND
potential $\f$ produced by a mass distribution $\r$ is a solution of
the Poisson equation for the modified source density
$\hr=-(\fpg)^{-1}\div\vg$, where $\vg=\n(|\vgN|/\az)\vgN$, and
$\vgN$ is the Newtonian acceleration field of $\r$. This makes $\f$
simply the scalar potential of the algebraic acceleration field
$\vg$. The theory thus involves solving only linear differential
equations, with one nonlinear, algebraic step. It is derivable from
an action, satisfies all the usual conservation laws, and gives the
correct center-of-mass acceleration to composite bodies. The theory
is akin in some respects to the nonlinear Poisson formulation of
Bekenstein and Milgrom, but it is different from it, and is
obviously easier to apply. The two theories are shown to emerge as
natural modifications of a Palatini-type formulation of Newtonian
gravity, and are members in a larger class of bi-potential theories.
\end{abstract}


\section{Introduction}
The only known, full-fledged, nonrelativistic (NR) formulation of
MOND has been that of Bekenstein \& Milgrom (1984). This is a
modified gravity theory in which the MOND gravitational potential,
$\f$, produced by a mass density $\r$, is gotten from the nonlinear
generalization of the Poisson equation:
 \beq \div[\m(|\gf|/\az)\gf]=4\pi G\r.  \eeqno{poisson}
Here $\m(x)$ is a function characterizing the theory.\footnote{There
are also the obvious generalizations to multi-potential theories,
whereby the MOND potential is a sum of several potentials, each
satisfying an equation such as eq.(\ref{poisson}).} This field
equation is derived from an action, and enjoys all the standard
conservation laws resulting from the usual space symmetries of the
underlying action.\footnote{Blanchet (2007) has given an
interpretation to this theory, as resulting from the omnipresence of
a gravitationally polarizable medium, and Blanchet \& Le Tiec (2008,
2009) extended the idea to an appropriate relativistic version.}
\par

\par
There is also extensive use in the literature of the pristine
formulation of MOND (Milgrom 1983), in which the MOND acceleration,
$\vg$, is calculated from the Newtonian value, $\vgN$, via an
algebraic relation of the form
 \beq \vg=\n(g\_N/\az)\vgN, \eeqno{algebraic}
 where $g\_N=|\vgN|$.
 For the case of circular motion in an
axisymmetric system, appropriate for rotation curve analysis,
eq.(\ref{algebraic}) is, in fact, an exact relation in the class of
theories dubbed ``modified inertia'' formulation of MOND (Milgrom
1994a). Expression (\ref{algebraic}) is also exact in the nonlinear
Poisson formulation for systems of one-dimensional symmetry, such as
a spherical galaxy. In this case $\n(y)$ is related to $\m(x)$ by
$\n(y)=1/\m(x)$ , where $x$ and $y$ are related by $x\m(x)=y$. This
algebraic formulation is very easy to use--hence its attraction as a
wieldy tool--and does capture the salient MOND effects in many
instances. However, it can definitely not be used as a
complete\footnote{`Complete' in the sense of being applicable to an
arbitrary problem in the realm for which it is meant; in the present
case, to any finite, self gravitating system.} theory, especially
for non-test-particle motion: In the first place $\vg$ resulting
from eq.(\ref{algebraic}) is, generally, not derivable from a
potential, with all the adverse effects of this (e.g., there is no
conserved momentum). Also, when applying eq.(\ref{algebraic}) to
describe the center-of-mass motion of a composite system based on
the accelerations of its constituents it fails completely. The same
is true when applying it, e.g., to the galactic external-field
effect in the solar system (Milgrom 2009a). This had left us with
only the nonlinear Poisson formulation of MOND as a reliable,
complete theory.
\par
Here I present a new MOND formulation that combines the benefits of
the above two formulations: It constitutes a complete theory
derivable from an action, and enjoying the standard conservation
laws. Yet, the unavoidable nonlinearity--a direct corollary of the
basic tenets of MOND--enters in an easy to handle, algebraic
manner. The resulting field equations then involve only linear
differential equations.
\par
This theory came to light as follows: In a recent paper (Milgrom
2009a), concerning the MOND external-field effect in the solar
system, I applied  the nonlinear Poisson formulation to a specific
problem: a point mass in a constant background field representing
the sun in the field of the Galaxy. Beside the exact treatment of
the problem, I considered a certain approximation for the MOND
potential of this configuration. It was defined as the solution of
the (linear) Poisson equation with a source density
$\hr=-(\fpg)^{-1}\div\vg$, with $\vg$ given by eq.(\ref{algebraic}).
This approximation was justified on the basis of $\hr$ having
properties similar  to those of the density-like field
$(\fpg)^{-1}\Delta\f$, where $\f$ is the exact solution of
eq.(\ref{poisson}). I have now come to realize that this
approximation may, in fact, form the basis for a complete MOND
theory, standing on its own.
\par
The nonlinear Poisson formulation of MOND has been applied,
analytically and numerically, to many problems to which it would be
interesting to apply the present theory as well. Among these: solar
system tests (Milgrom 1986a, 2009a, Bekenstein \& Magueijo 2006),
forces on non-test-mass bodies (Milgrom 1997,2002a, Dai Matsuo \&
Starkman 2008), disc stability and bar formation (Brada \& Milgrom
1999, Tiret \& Combes 2008), two-body relaxation (Ciotti \& Binney
2004), dynamical friction (Nipoti \& al. 2008), escape speed from a
galaxy (Famaey, Bruneton, \& Zhao 2007, Wu et al. 2007), galaxy
interactions (Tiret \& Combes 2007, Nipoti Londrillo \& Ciotti
2007a), and collapse (Nipoti, Londrillo \& Ciotti 2007b), triaxial
models of galactic systems (Wang, Wu, \& Zhao 2008, Wu \& al. 2009),
the external-field effect as applied to dwarf spheroidals and warp
induction (Brada \& Milgrom 2000a,b, Angus 2008), structure
formation (e.g., Llinares, Knebe, \& Zhao 2008), and quite a few
more.
\par
Relativistic extensions of the nonlinear Poisson theory where also
propounded (see, e.g., Sanders 1997, Bekenstein 2004, 2006, Zlosnik
Ferreira \& Starkman 2006, 2007, Skordis 2009). The bi-, or
multi-potential nonrelativistic theories described here have also
inspired a class of relativistic, bi-, or multi-metric MOND theories
(Milgrom 2009c,d, see below).
\par
The new, quasi-linear MOND (QUMOND) theory--in particular its
emergence as a modification of a Palatini formulation of Newtonian
gravity, and its generalizations--is described in section
\ref{theory}. In section \ref{forces} I derive some of its general
properties pertaining to forces on bodies. Section \ref{deepmond}
concern the deep-MOND limit of the theory. Section \ref{constant}
deals with systems in a constant external field, relevant for the
external field effect, and for the question of the center of mass
motion of composite systems. Section \ref{discussion} is a
discussion.

\section{Formalism}
\label{theory} Consider a gravitating system of density $\r(\vr,t)$
and velocity field $\vv(\vr,t)$. Describe the dynamics of the system
by the action $I=\int L dt$, with $L\equiv \int\L \drt $, and
 \beq \L=\L\_K-\L\_P=-{1\over 8\pi
G}\{2\gf\cdot\gfN-\az^2\Q[(\gfN/\az)^2]\}
  +\r({1\over 2}\vv^2-\f),  \eeqno{iii}
  involving the two
potentials $\f$ and $\fN$, where $\L\_K =\r\vv^2/2$ is the kinetic
energy density, and $\L\_P$ is the potential energy density. The
density $\r$ may be viewed as made up of the masses of the
constituents $\r(\vr,t)=\sum_i m_i\d^3[\vr-\vr_i(t)]$, with each of
the masses, $m_i$, negligible with respect to the total mass; so
each mass can be treated as a test mass in the field of the rest.
\par
Varying the action over the particle degrees of freedom gives
 \beq \ddot\vr_i=-\gf(\vr_i).  \eeqno{mot}
 Varying over $\f$ gives
 \beq \Delta\fN=\fpg\r,
 \eeqno{pois}
 and varying over $\fN$ gives
 \beq \Delta\f=\div[\n(|\gfN|/\az)\gfN],
 \eeqno{poisa}
where $\n(y)\equiv \Q'(y^2)$. Equation (\ref{mot}) tells us that the
masses move according to the standard Newtonian law of inertia in
the potential $\f$; thus $\f$ is the MOND potential. Equation
(\ref{pois}) tells us that $\fN$ solves the Poisson equation with
$\r$ as source; thus $\fN$ is the standard Newtonian potential (when
we impose the standard boundary condition $\fN\rar 0$ at infinity).
Equation (\ref{poisa}) tells us that the MOND potential satisfies
the Poisson equation for the density
 \beq
 \hr=-{1\over\fpg}\div[\n(g\_N/\az)\vgN]=
 \n\r-{1\over\fpg\az}\n'\grad g\_N\cdot\vgN
 \eeqno{phua}
as a source, with $\vgN=-\grad\fN$. Thus $\hr$ would be the density
that gives the correct MOND potential if we interpret the
observations in the framework of Newtonian dynamics. Then,
$\rp\equiv\hr-\r$ would be the phantom mass density, introduced in
Milgrom (1986b), which will be interpreted by a Newtonist as the
density of dark matter. The density $\rp$ is what was called in
Milgrom (2009a) ``the surrogate mass density'', and was used as an
approximation for the phantom density of the nonlinear Poisson
theory.
\par
As usual, the symmetries of the action under space and time
translations, and rotations,  lead to conserved momentum, energy,
and angular momentum (see also section \ref{forces}).
\par
In effect, this theory starts with the acceleration field defined by
the algebraic relation eq.(\ref{algebraic}), and remedies its not
being derived from a potential by projecting it on the space of
gradient vector fields. In other words, write
 \beq \n(g\_N/\az)\vgN=-\gf-\curl\vA,  \eeqno{vure}
which is a unique decomposition if we require that $\gf$ vanish at
infinity, then $\f$ is the MOND potential of the theory.
\par
Restoration of the Newtonian limit for $\az\rar 0$ requires
$\Q(z)\rar z+Q_1$ for large $z$, where $Q_1$ is a constant.
Space-time scale invariance, which is the defining tenet of the
deep-MOND limit $\az\rar\infty$ (Milgrom 2009b), and the standard
normalization of $\az$ (defined so that the
mass-asymptotic-rotational-speed-relation is $V^4=MG\az$) for which
$\n(y)\approx y^{-1/2}$, dictate $\Q(z)\approx (4/3)z^{3/4}+Q_0$
for $z\ll 1$. Since $Q$ is defined up to an immaterial additive
constant, we choose it so that $Q_0=0$.
\par
In one dimensional cases--e.g., for spherically symmetric
systems,--eq.(\ref{poisa}) implies relation (\ref{algebraic}), and
all three formulations give the same acceleration field with
$\n(y)=1/\m(x)$, where  $y=x\m(x)$. The QUMOND theory will produce
unique solutions if and only if $y\n(y)$ is monotonic, which we
assume; and since this function is $\approx y$ for large values of
$y$, it has to be increasing everywhere, so $y\n'+\n>0$.
\par
For an isolated mass distribution of total mass $M$, bounded in a
finite region, we have asymptotically at infinity $\fN\approx
-MG/r$. Thus $\hr\approx
(\fpg)^{-1}(MG\az)^{1/2}r^{-2}=(1/3)\r\_M(r/\RM)^{-2}$, where
$\r\_M=3M/4\pi\RM^3$, $\RM=(MG/\az)^{1/2}$, are, respectively, the
MOND density and the MOND radius for the mass $M$. It thus follows
that the MOND potential, relative to its value at a finite point,
diverges logarithmically at infinity (as in any MOND theory). Note
also that the above asymptotic behavior of the fields makes $\LP$
logarithmically divergent upon space integration. This is the same
situation as in the case of the nonlinear Poisson formulation.
Differences in $\LP$ for systems with the same total mass are
finite, and only such differences will concern us.
\par
We can write the MOND potential, relative to some arbitrary origin
$\vr=0$, as
 \beq \f(\vr)=-G\int d^3r'\hr(\vr')
 \left({1\over |\vr-\vr'|}-{1\over |\vr'|}   \right)=
 {1\over 4\pi}\int d^3r'\div[\n(g\_N/\az)\vgN](\vr')
 \left({1\over |\vr-\vr'|}-{1\over |\vr'|}   \right),
\eeqno{loit} which is finite. This is a closed form expression of
the MOND field for an arbitrary (bounded) mass distribution, with
$\vgN$ itself being expressed as usual in terms of $\r$.
\subsection{Emergence from a Palatini formulation}
The proposed modification amounts to replacing $(\gf)^2$ in the
standard Newtonian Lagrangian density
 \beq \L=-{1\over 8\pi G}(\gf)^2 +\r({1\over 2}\vv^2-\f)
  \eeqno{iiiga}
 by $2\gf\cdot\gft-\az^2\Q[(\gft/\az)^2]$, where $\ft$ is an
auxiliary potential, called here $\fN$ because it turns out to equal
the Newtonian potential. This modification can be performed in two
steps: First add to $\L$ a term $(8\pi G)^{-1}(\gf-\gft)^2$, which
is the same as $-(8\pi G)^{-1}(\f-\ft)(\Delta\f-\Delta\ft)$ up to a
divergence. This does not modify the theory since varying over
$\ft$ gives $\Delta\ft=\Delta\f$, which annuls the new addition to
the action. The free field part of the new Lagrangian density now
has $-(\gft)^2+2\gf\cdot\gft$ replacing $(\gf)^2$.
\par
The resulting theory after this step may be viewed as a
Palatini-type formulation of Newtonian gravity. In the standard
Palatini formulation of general relativity (GR), the connection
(representing an auxiliary gravitational acceleration field) is not
assumed, a priori, to be the Levi-Civita connection of the metric.
The analogous approach here would be to introduce an
auxiliary-acceleration degree of freedom, $\vgt$, and not assume, a
priori, that it is the gradient of the potential. Write then the
Lagrangian density as
 \beq \L={1\over 8\pi
G}(\vgt^2-2\f\div\vgt)
  +\r({1\over 2}\vv^2-\f).  \eeqno{iiisut}
Variation over $\vgt$ gives $\vgt=-\gf$, and over $\f$,
$\div\vgt=-\fpg\r$, yielding Newtonian gravity.\footnote{The second
term in the first part of $\L$ can be replaced by $+2\gf\cdot\vgt$
from which it differs by a divergence.} The first part in $\L$ falls
from the gravitational action $-c^4(16\pi G)^{-1}\int\gh\Gmn\tilde
R\ten{}{\m\n}$ of the relativistic Palatini formulation of GR. Then,
$\vgt$ descends from the independent connection degrees of freedom,
from which $\tilde R\ten{}{\m\n}$ is constructed, and $\f$ from the
metric. The second part in $\L$ falls from the matter Lagrangian
density, with matter coupled minimally to the metric, and not to the
connection. In the relativistic Palatini formulation, extremization
over the connection (assumed symmetric) establishes it as the
Levi-Civita connection of the metric (here, $\vgt$ is established as
the gradient of the gravitational potential), and extremization over
the metric (here, $\f$) gives the Einstein equation for the
connection with the standard matter energy-momentum as source (here,
the Poisson equation for $\f$).
\par
MOND is then introduced in the second step, replacing $\vgt^2$ in
$\L$ of eq.(\ref{iiisut}) by $\az^2Q(\vgt^2/\az^2)$. To get our
Lagrangian (\ref{iii}), we now do impose, beyond the standard
Palatini formalism, that $\vgt$ is a gradient of some auxiliary
potential (not of $\f$ itself): $\vgt=-\gft$; so, $\ft$, not $\vgt$,
is the basic degree of freedom beside $\f$. If we do not impose this
restriction on $\vgt$ we get instead of
eqs.(\ref{pois})(\ref{poisa}) a different theory
 \beq \nu(|\vgt|/\az)\vgt=-\gf,~~~~~\div\vgt=-\fpg\r,\eeqno{veiop}
 which, quite interestingly, is equivalent to the nonlinear
Poisson formulation, eq.(\ref{poisson}), in the form used, e.g., in
Milgrom (1986a) [with $\vgt=-\m(|\gf|/\az)\gf$].
\par
To recapitulate, the Lagrangian density (\ref{iiisut}) underlies a
Palatini-like formulation of Newtonian gravity, whether or not we
restrict $\vgt$, a priori, to be a gradient. However, the MOND
extension, even with a given $\Q(z)$, does depend on whether we make
the restriction or not: With the restriction, we get our present
QUMOND theory. Without it--varying the action over a larger space of
trial acceleration fields $\vgt$--we get the nonlinear Poisson
theory. This captures at once the affinity and the disparity between
the two theories, and also explains why they coincide in cases of
one-dimensional symmetry, where any vector field is a gradient.
\subsection{Generalizations}
\label{generalizations} Since the Palatini formulation introduces a
second degree of freedom, we can further generalize the Lagrangian
density\footnote{We require invariance under $\f\rar \f+const$. Also
we limit ourselves here to actions where only $\vgt$ appears, and
not its derivative, because we want to compare with theories where
$\vgt$ is a gradient of some potential, and we want to avoid higher
derivatives of that potential.} (\ref{iiisut}) by taking
 \beq \L\_P=-{\az^2\over 8\pi G}\F(\z,\k,\xi)-\r\f,~~~{\rm with}~~
\z=(\vgt/\az)^2,~\k=(\gf/\az)^2,~\xi=2\vgt\cdot\gf/\az^2.
\eeqno{kurte}
 Extremization over $\vgt$ and over $\f$ gives, respectively
 \beq \F_{\z}\vgt=-\F_{\xi}\gf, \eeqno{bupe}
 \beq \div(\F_{\k}\gf+\F_{\xi}\vgt)=\fpg\r.  \eeqno{hutt}
Since by eq.(\ref{bupe}) $\vgt$ and $\gf$ must be aligned, one of
the variables, e.g. $\k$, determines the other two, e.g., by solving
the equations $\F_{\z}^2\z=\F_{\xi}^2\k=-\F_{\z}\F_{\xi}\xi/2$,
which follow from eq.(\ref{bupe}). Substituting the resulting
expressions in eq.(\ref{hutt}) then gives
$\div[\m(\k^{1/2})\gf]=\fpg\r$. So even in this general case we get
the nonlinear Poisson theory eq.(\ref{poisson}), and $\F(\z,\k,\xi)$
enters the dynamics of the system only through the function $\m$ of
a single variable, extracted from it.
\par
If, however, we assume a priori that $\vgt$ is a gradient field
$\vgt=-\gft$, the variety of choices of $\F$ in the above Lagrangian
yields a richer family of theories
 \beq \div(\F_{\z}\gft-\F_{\xi}\gf)=0,~~~~
\div(\F_{\k}\gf-\F_{\xi}\gft)=\fpg\r. \eeqno{metar} It includes our
QUMOND theory with $\F_{\z}=-\n(\z^{1/2}),
 ~\F_{\k}=0,
~\F_{\xi}=-1$, and also the nonlinear Poisson theory with
$\F_{\z}=\F_{\xi}=0,~~\F_{\k}=\m(\k^{1/2})$.
\par
Another interesting subclass is
 \beq \F=\O(\z)+\N(\k)-\M(\z+\k+\xi)   \eeqno{cuit}
[note that $\z+\k+\xi=(\gf-\gft)^2/\az^2$]. With $\O=0$ and $\N=\k$
we get Newtonian dynamics. For $\N=0$ we have
$\F_{\k}=\F_{\xi}=-\M',~\F_{\z}=\O'-\M'$; so, defining $\fs=\f-\ft$,
we get from eq.(\ref{metar})
 \beq
 \div\{\O'[(\gft/\az)^2]\gft\}=
 -\div\{\M'[(\gfs/\az)^2]\gfs\}=\fpg\r.
 \eeqno{kurtas}
Thus, $\f=\ft+\fs$, with $\ft$ and $\fs$ satisfying the nonlinear
Poisson equation. This covers the NR limit of TeVeS for $\O\propto
\z$, so $\ft$ satisfies the linear poisson equation.
\par
An interesting subfamily is $\O=\a\z~(\a\not =0),~\N=\b\k$, for
which the Lagrangian density is
  \beq \L=-{1\over 8\pi
G}\{\a(\gft)^2+\b(\gf)^2-\az^2\M[(\gf-\gft)^2/\az^2]\}
  +\r({1\over 2}\vv^2-\f),  \eeqno{futcol}
or, in terms of $\f$ and $\fs=\f-\ft$:
  \beq \L=-{1\over 8\pi
G}\{(\a+\b)(\gf)^2-2\a\gf\cdot\gfs+\a(\gfs)^2
-\az^2\M[(\gfs)^2/\az^2]\}
  +\r({1\over 2}\vv^2-\f),  \eeqno{futram}
where I used the freedom to normalize $\a,~\b$, and $\M$ so that $G$
is the Newton constant. The field equations are then
$$\div[\m^*(|\gfs|/\az)\gfs]=\fpg\r,$$
  \beq \D\f=\fpg\b^{-1}\r+\b^{-1}\div(\M'\gf^*)
  =\div[(1-\a^{-1}\M')\gfs],
\eeqno{hutred}
  with
   \beq \m^*=\b-{\a+\b\over\a}\M'[(\gfs/\az)^2]. \eeqno{cutes}
One has to solve the nonlinear Poisson equation for $\fs$, and then
the linear version for $\f$. In fact, for $\a+\b\not= 0$,
the second eq.(\ref{hutred}) can be written as
 \beq \D\f={1\over \a+\b}\D(\a\fs+\fN), \eeqno{kubas}
so the MOND potential is simply a linear combination of the
Newtonian potential and a solution of the first of
eq.(\ref{hutred}), which is of the type of eq.(\ref{poisson}).
\par
The case $\a+\b=0$ ($\m^*=\b$) is an
interesting special case as it gives our QUMOND theory (see below).
\par
Consider now the Newtonian and MOND limits of the theory underlaid
by the Lagrangian density (\ref{futram}). It is enough to consider
the spherical case so that applying Gauss theorem to the field
equations (\ref{hutred}) we get algebraic relations between the
(minus) radial accelerations $g=d\f/dr\ge 0,~g^*=d\fs/dr$, and the
Newtonian acceleration $g\_N=d\f\_N/dr\ge 0$:
 \beq \m^* g^*=g\_N,~~~~~ g=(1-\a^{-1}\M')g^*={\a-\M'\over \a\b
 -(\a+\b)\M'}g\_N.\eeqno{vyter}
In the Newtonian limit $g/g\_N\rar 1$ so
 \beq \M'\rar \M'\_{\infty}={\a(\b-1)\over \a+\b-1}. \eeqno{gersa}
This relation between $\a,~\b,$ and $\M'\_{\infty}$ does not require
any tuning between different contributions to the action. It
reflects our taking $G$ to be the phenomenological Newton constant.
Starting with a coupling $G'$, we would end up with a relation
$\M'\_{\infty}=\a(\b-G'/G)/( \a+\b-G'/G)$, which is only used to
define Newton's constant in terms of $\a,~\b$, $G'$, and
$\M'\_{\infty}$.
\par
If $\M$ is such that $\M'\_{\infty}=0$, we must have $\b=1$ to get
the correct Newtonian limit (or rather $G=G'/\b$, which allows us to
normalize the coefficients so that $G'=G,~\b=1$), $\a$ is not
constrained. For $\M'\_{\infty}$ infinite we have to have $\a+\b=1$.
 When $\M'\_{\infty}$ is finite we have neither $\b=1$ nor
 $\a+\b=1$.
\par
We can write $g$ as a function of $g\_N$ alone, distinguishing
between two cases. If $\a+\b\not = 0$
 \beq g^*=\n^*(|g\_N|/\az)g\_N,~~~~g=g\_N
 +{1-\a-\b\over \a+\b}g\_N+{\a\over \a+\b}\n^*(|g\_N|/\az)g\_N,
\eeqno{nitrir} where $\n^*(y)$, as before, is such that if
$y=x\m^*(x)$ then $x=y\n^*(y)$. If $\a+\b=0$ we have
 \beq g^*=\b^{-1}g\_N,~~~~~g=g\_N+\b^{-2}\{\M'[(g\_N/\b\az)^2]
 -\M'\_{\infty} \}g\_N,
 \eeqno{nirfas}
where here, according to eq.(\ref{gersa}), $\M'\_{\infty}=\b(\b-1)$.
From these we can read the requirements for the Newtonian and MOND
limits. Consider first the case $\a+\b\not =0$. In the MOND limit,
$g\_N/\az\rar 0$ we have to have $g/g\_N\rar (g\_N/\az)^{-1/2}$
diverging. The last term in expression (\ref{nitrir}) dominates, and
we must have $[\a/(\a+\b)]\n^*(y)y\rar y^{1/2}$.
\par
In considering the Newtonian limit for the $\a+\b\not = 0$ case, we
have to fork again: If $\a+\b=1$ (which, as we saw, applies when
$\M'\_{\infty}$ is infinite) the second term in the second of
equations (\ref{nitrir}) vanishes, and so $\n(y)y$ has to vanish for
$y\rar\infty$. However, if the first of eqs.(\ref{hutred}) is to be
elliptic, as it must, $y\n^*(y)$ must be monotonic function. We saw
that in the MOND regime $[\a/(\a+\b)]\n^*(y)y\approx y^{1/2}$ is an
increasing function, and so it must be increasing everywhere. This
means that $g/g\_N-1$ must vanish in the Newtonian limit slower than
$(g\_N/\az)^{-1}$. This, however, is in conflict with solar system
constraints (Milgrom 1983, 2009a, Sereno \& Jetzer 2006); this would
thus rule out the case $\a+\b=1$.\footnote{This finding is of the
same kind arrived at by Zhao and Famaey (2006) in regard to certain
versions of TeVeS.}
\par
When $\a+\b\not =1$, $\n^*$ has to tend to a constant value for
large arguments: $\n^*(y)\rar (\a+\b-1)/\a$, and this it can do with
arbitrary speed without violating the ellipticity condition.
However, comparing the MOND and Newtonian limits in this case we see
that if $0<\a+\b<1$, $\n^*$ has to vanish for some finite value of
its argument, and this means that $\m^*(x)=1/\n^*(y)$ as to blow up
at a finite argument value. This is undesirable, and excludes the
parameter range $0<\a+\b<1$, leaving us with  $\a+\b>1$ or $\a+\b\le
0$. There may be other constraints on $\a,~\b$ to be investigated
(e.g., positivity conditions).
\par
Now return to the interesting boundary case $\a+\b=0$. There is no
matter-of-principle constraint on how fast $\M'-\M'\_{\infty}$ can
vanish for large arguments; so this theory can be made to approach
Newtonian dynamics arbitrarily fast. The MOND limit phenomenology
dictates $\M'(z)\approx \b^{3/2}z^{-1/4}$ for $z\ll 1$. In fact, we
can, when $\a+\b=0$, absorb $\b$ in the definition of $\fs$, so that
$\fs\rar \b^{-1}\fs$, giving an equivalent theory with
$\M(z/\b^2)-(1-\b^{-1})z$ as the new $\M$. The new $\M'$ then
vanishes for high arguments. Without loss of generality we can thus
put in this case $\b=1$ and $\M'\_{\infty}=0$, with the equations of
motion
   \beq \D\fs=\fpg\r,
~~~~~~\D\f=\div[(1+\M')\gfs]=\fpg\r+\div(\M'\gfs). \eeqno{hutram}
  This theory is
equivalent to the QUMOND theory I started with, with
$\Q(z)=z+\M(z)$, and so $\n(y)=1+\M'(y^2)$. $\gfs$ is the Newtonian
acceleration, and $(\fpg)^{-1}\div(\M'\gfs)$ is the density of the
``phantom matter'' representing DM. In the MOND limit $z\rar 0$ we
have $\M'(z)\rar z^{-1/4}$.
\par
In all the above examples the theory can be cast as two equations
(for two potentials) that can be solved separately [as for the
system (\ref{kurtas})], or sequentially [as for system
(\ref{hutred})]. For a general $\F$ this is presumably not possible
and solving the resulting theory is rather more challenging.
\par
Generalizing even further, we could include several auxiliary
acceleration fields, $\vg_i$, in addition to $\f$, and have $\L\_P$
a function of all the scalars
$\vg_i^2,~(\gf)^2,~\vg_i\cdot\vg_j,~\vg_i\cdot\gf$. Without
constraining $\vg_i$ to be gradients we still get the nonlinear
Poisson theory; with the constraints we seem to get a yet richer
family. All the above theories are equivalent for a spherical
system.
\subsection{Extensions}
Here I describe succinctly two interesting extensions of the class
of theories described above.
\par
These bi-potential theories have inspired a class of relativistic
formulations of bi-metric MOND (BIMOND) theories (Milgrom 2009c,d).
The BIMOND theories not only were constructed in analogy with the
bi-potential theories discussed here, but, in fact, reduce to them
in the NR limit. The Lagrangian density of the BIMOND theories is
constructed after the fashion of the Lagrangian density
(\ref{futcol}): Instead of two potentials we now have two metrics.
One, $\gmn$, is the MOND metric, descending to the MOND potential in
the NR limit, and the other, $\hgmn$, is an auxiliary one giving
$\ft$ in the limit. Matter couples only to the MOND metric in the
standard way, echoing the fact that here matter couples only to
$\f$. As in many bimetric formulations discussed in the literature,
the free Lagrangian densities $(\gf)^2, ~(\gft)^2$ are replaced by
the corresponding Ricci scalars of the two metrics. The novelty
enters in designing the interaction term between the two metrics.
The difference between the Levi-Civita connections of the two
metrics,
 \beq  C\ten{\a}{\b\c}=\C\ten{\a}{\b\c}-\hC\ten{\a}{\b\c},
  \eeqno{veyo}
is a tensor, which, moreover, plays the role of gravitational
accelerations. Such acceleration-like tensors are obviously crucial
in the context of MOND, as they permit us to construct dimensionless
scalars from $\az^{-1}C\ten{\a}{\b\c}$ and then take functions of
these to serve as interpolating functions between the GR and the
MOND regime.\footnote{Here $c=1$, otherwise we use the MOND
scale-length $\ell=c^2/\az$ in the dimensionless tensors $\ell
C\ten{\a}{\b\c}$.}
\par
I have thus considered actions of the form
   \beq I=-{1\over 16\pi G}\int[\b\gh R + \a\hgh \hat R
 -(g\hat g)^{1/4}f(\k)\az^2\M(\Up/\az^2)]d^4x
 +I\_M(\gmn,\psi_i)+\hat I\_M(\hgmn,\chi_i),  \eeqno{gedap}
where in the argument of $\M$, $\Up$ is a scalar quadratic in the
tensor $C\ten{\a}{\b\c}$. As in our NR case $\M'$ plays the role of
an interpolating function between the MOND and GR regimes. $I\_M$ is
the matter action, with matter degrees of freedom represented by
$\psi_i$, coupling only to $\gmn$ [$g$ and $\hat g$ are minus the
determinants of the two metrics, and $\k=(g/\hat g)^{1/4}$]. I also
allow for `twin' matter described by degrees of freedom $\chi_i$,
which couples only to $\hgmn$. Since the scalar $\Up$ contains only
first derivatives of the metrics this action leads to second order
field equations.
\par
The other interesting extension, inspired, in return, by the BIMOND
theories, is the possible inclusion of `twin' matter in our NR
theories. This can be effected by adding to our Lagrangian density
(\ref{futcol}) a term $\c\tilde\r({1\over 2}\tilde\vv^2-\ft)$ (which
would be the NR limit of $\hat I\_M$). We then have two types of
matter, each accelerated directly, in the standard manner, only by
its own potential, but interacting indirectly, `gravitataionally',
through the coupling between their potentials, rather unlike
standard gravity. The field equations (\ref{hutred}) now read
$$\div[\m^*(|\gfs|/\az)\gfs]=\fpg(\r-\b\c\a^{-1}\tilde\r),$$
  \beq \D\f=\fpg\b^{-1}\r+\b^{-1}\div(\M'\gf^*)
  =\div[(1-\a^{-1}\M')\gfs]+\fpg\c\a^{-1}\tilde\r.
\eeqno{hugtal} I will discuss the interesting implications of such
theories with twin matter in a separate paper (Milgrom, in
preparation), taking in the rest of this paper $\tilde\r=0$.

\section{Forces and virial relations}
\label{forces}  I concentrate hereafter on the simpler, QUMOND
special case. Most results are easily carried to the more general
case. Because of the similarities between the theories, many of the
properties of the nonlinear Poisson formulation derived in the past
(see e.g., Bekenstein \& Milgrom 1984, Milgrom 1986a,b, 1994b, 1997,
1998, 2002a) invite derivation of analogous properties of the QUMOND
formulation. I derive some of these here and in the following
sections.
\subsection{Forces}
The force $\vF$ on the collection of masses in a volume $V$ is
defined as the generator of translations, in the sense that under a
rigid translation of the masses in $V$ alone by a small increment
$\d\vr$ the potential energy changes by
 \beq \d \LP=-\d\vr\cdot\vF.   \eeqno{gutet}
Such a translation causes a change in the density
$\d\r=-\d\vr\cdot\grad\r$, inside $V$ and zero outside. This
generates corresponding changes in the potentials; but taking into
account the stationarity of $\LP$ under such changes (since the
total mass is fixed, the variation in the potentials vanishes at
infinity) we have $\d \LP=\int_V\f\d\r=-\d\vr\cdot\int_V\f\grad\r$.
Integrating by parts then gives
 \beq \vF=-\int_V\drt\r\gf.  \eeqno{force}
\par
It is useful to consider the stress tensor of the gravitational
field, $\TP$. One way to drive it is to write $\LP$ as a coordinate
scalar in curved space, and consider its variation under a change
$\d\gij$ in the metric, to get $\TP$ from
 \beq \d \LP={1\over 2}\int g^{1/2}\drt \TP_{ij}\d\Gij,  \eeqno{molita}
where summation over repeated indices is understood. One finds from
eq.(\ref{iii}), going back to the Euclidean case,
 \beq \fpg \TP=(\vg\cdot\vgN-{\az^2\over 2}\Q){\rm I}+\n\vgN\otimes\vgN
 -\vg\otimes\vgN-\vgN\otimes\vg, \eeqno{kitreq}
where ${\rm I}$ is the unit tensor.\footnote{Since $\TP$ is derived
from the action without reference to whether $\vg_N$ is a gradient,
this expression for $\TP$, and the subsequent expressions that use
it, remains valid in the nonlinear Poisson formulation if we replace
$\vgN$ by $-\m\gf$, and $\vg$ by $-\gf$. This gives $\fpg
\TP=[\m(\gf)^2-{\az^2\over 2}\Q]{\rm I}-\m\gf\otimes\gf$, which
agrees with the expression given in Milgrom (2002a).} For solutions
of the equations of motion, the divergence of $\TP$ is found to be
 \beq \div\TP= \r\vg.  \eeqno{lovy}
We can thus write the force on $V$ as an integral over its surface
$\S$:
  \beq \vF=\int_V\drt\r\vg=\int\_{\S} \TP\cdot\vds, \eeqno{opio}
 or
  \beq \vF={1\over \fpg}\int\_{\S}(\n\vgN-\vg)\vgN\cdot\vds-
  \vgN\vg\cdot\vds+(\vg\cdot\vgN
  -{\az^2\over 2} \Q)\vds.  \eeqno{juipo}
The torque on the volume $V$--the generator of rotations in the
above sense--is

\beq \vT=\int_V\drt\r\vr\times\vg=\int\_{\S} \vr\times\TP\cdot\vds,
\eeqno{opiona} where I integrated by parts making use of the
symmetry of $\TP$,
 or
  \beq \vT={1\over \fpg}\int\_{\S}[\vr\times(\n\vgN-\vg)]\vgN\cdot\vds
  -\vr\times\vgN\vg\cdot\vds
  +(\vg\cdot\vgN-{\az^2\over 2}\Q)\vr\times\vds.  \eeqno{juipom}
In Appendix \ref{A} I give other expressions for $\vF$ and $\vT$ as
surface integrals, which might also be useful.
\par
For an isolated bounded system we get vanishing total force and
torque, as can be seen by taking the integration surface in
eqs.(\ref{juipo})(\ref{juipom}) as the sphere at infinity, and using
the asymptotic behavior of the fields. This is tantamount to the
conservation of the total momentum and angular momentum in an
isolated system.

\subsection{Virial relations}
Multiplying eq.(\ref{poisa}) by $\fN$ and integrating over all space
gives an integral relation satisfied by solutions of the field
equations:\footnote{With the MOND behavior of $\n$ and the
asymptotic fields, each of the two terms in the integrand gives rise
to an integral that diverges logarithmically; the integral of the
difference not only converges, but vanishes.}
 \beq \int[\vgN\cdot\vg- \n(g\_N/\az)g\_N^2]\drt=0.
 \eeqno{nurteq}
This relation, together with the useful inequality
 \beq zQ'(z)>{1\over 2}Q(z),~~~~~{\rm for} ~~~ z>0,  \eeqno{curteq}
implies that the free-field energy $(8\pi G)^{-1}
\int\drt[2\gf\cdot\gfN-\az^2\Q]$ is always positive for solutions of
the field equations.\footnote{The energy integral actually diverges
logarithmically for an isolated system.} To derive inequality
(\ref{curteq}) look at $H(y)\equiv zQ'(z)-Q(z)/2=y^2\n(y)-Q(y^2)/2$,
with $z=y^2$. We have $H(0)=0$ because $\Q(0)=0$, and
$H'(y)=y(y\n'+\n)>0$ for $y>0$, from the uniqueness condition; so
$H(y)>0$ for $y>0$.
\par
Define the virial as
  \beq \V\equiv -\int\drt\r\vr\cdot\vg, \eeqno{viva}
which is useful in several contexts. For example, since the force
density on a system can be written as  $\vf(\vr)=\r(\vr)\vg(\vr)$,
we can write $\V=-\int\drt\vr\cdot\vf$. The virial is the generator
of space scaling in the sense that under $\vr\rar(1+\eps)\vr$ for
small $\eps$, $\LP\rar \LP+\eps\V$ (see detailed discussion in
Milgrom 1997).
 Using eq.(\ref{lovy})
we write
 \beq
 \V=-\int\drt(\div\TP)\cdot\vr=-\int\drt\div(\TP\cdot\vr)+\int\drt\P,
 \eeqno{huttter}
where the trace
 \beq \P\equiv {\rm Tr}(\TP)={1\over
\fpg}[\n g\_N^2+\vg\cdot\vgN-{3\over
 2}\az^2\Q],\eeqno{hiop}
Write the first term in eq.(\ref{huttter}) as a surface integral to
get
 $$\fpg\V=\int\drt [2\n g\_N^2-{3\over
 2}\az^2\Q]-$$
 \beq -\int\_{\S}(\vg\cdot\vgN-{\az^2\over 2}\Q)\vr\cdot\vds
+\n\vr\cdot\vgN\vgN\cdot\vds-
\vr\cdot\vg\vgN\cdot\vds-\vr\cdot\vgN\vg\cdot\vds , \eeqno{moret}
where I made use of the integral relation
(\ref{nurteq}).\footnote{Each of the terms in expression
(\ref{hiop}) for $\P$ behaves as $r^{-3}$ at infinity, but $\P$
itself vanishes faster, so it gives a finite integral over
space.}$^,$\footnote{The above expressions for $\vF,~\vT$, and $\V$
are invariant to an addition of a constant to $\Q$, as they should
be.} In Appendix \ref{A} I give another useful expression for the
virial.
\par
In a theory in which $\Q(z)$ vanishes at small $z$ faster than the
MOND behavior $\Q\propto z^{3/4}$, such as in Newtonian dynamics,
the surface integrals vanish at infinity, and what remains of
eq.(\ref{moret}) forms the basis of a virial relation upon its
integration over time. In the case of MOND both the volume integral
and the surface integrals are finite. Integrating over a sphere at
infinity, all the vectors appearing in the surface integral in
eq.(\ref{moret}) are radial. It follows then that all cancel except
$(\az^2/2)\int\_{\S}\Q\vr\cdot\vds$, which contributes $(8\pi
G/3)M(MG\az)^{1/2}$, where $M$ is the total mass of the
system.\footnote{This applies for the particular choice of the
additive constant in the definition of $\Q$ for which $\Q(0)=0$.}
So,
 \beq \V={1\over \fpg}\int\drt[2\n g^2\_N-{3\over
 2}\az^2\Q]+{2\over 3}M(MG\az)^{1/2}.  \eeqno{kiop}
\par
If we view $\r$ as made up of many discrete small point masses, so
that each can be considered a test particle in the field of the
rest, then the force, $\vF_i$, on particle $i$ at position $\vr_i$
is $-m_i\gf(\vr_i)$ (this is not true for a non-test particle). We
can then write the virial as\footnote{In the fluid approximation,
use of the Euler and continuity equations leads directly to
$\V=\int\drt\r\vv^2-(d/dt)\int\drt\r\vr\cdot\vv$.}
  \beq \V=-\sum_i\vr_i\cdot\vF_i=-\sum_i m_i\vr_i\cdot\ddot\vr_i=
  \sum_i m_i\vv^2_i-{d\over dt}\sum_i m_i\vr_i\cdot\vv_i.
\eeqno{fepot}
 Use eq.(\ref{kiop}) and, as usual, average over a long time,
 dropping the last term in eq.(\ref{fepot}) for systems that remain
 bound over long times, we get for such systems the virial relation
 \beq\langle\vv^2\rangle\equiv
 M^{-1}\langle\sum_i m_i\vv^2_i\rangle_t
={2\over 3}(MG\az)^{1/2}+{1\over 4\pi MG}\langle\int\drt[2\n
g^2\_N-{3\over
 2}\az^2\Q]\rangle_t, \eeqno{virqet}
where $\langle\rangle_t$ denotes the long-time average.
\par
There is an analogous relation that holds in the nonlinear Poisson
formulation (Milgrom 1994b), but the integral there involves the
MOND potential, for which we have to solve before we can use this
virial relation. In the present theory the integral involves only
the Newtonian acceleration field and is readily calculated from the
mass distribution--an example of the added amenability of the
present formulation.

\section{The deep-MOND limit}
\label{deepmond}
 In the deep-MOND limit the Lagrangian
density becomes
 \beq
 \L={\az^2\over 6\pi G}[(\gfN/\az)^2]^{3/4}
 -{1\over \fpg}\gf\cdot\gfN +\r({1\over 2}\vv^2-\f).
\eeqno{ivv} Multiplying by $\az G$ and absorbing $\az\fN\rar\fN$
gives an  action with only $\az G m_i$ appearing, as required by the
basic MOND tenets. Seen differently, under space-time scaling
$(\vr,t)\rar\l(\vr,t)$, with $\f$ having zero internal dimension so
that $\f(\vr)\rar \f(\vr/\l)$, and $\fN$ having dimension $-1$, so
$\fN(\vr)\rar \l^{-1}\fN(\vr/\l)$, $L$ is invariant. The action is
thus multiplied by $\l$; so the equations of motion of the masses
are invariant. This is the defining requirement of the deep-MOND
limit of any theory (Milgrom 2009b).
\par
For the present theory there exists another symmetry of the
potential Lagrangian, $\LP$, alone: Under space scaling
$\vr\rar\l\vr$, for a constant $\l$, where, again,
$\f(\vr)\rar\f(\vr/\l)$, $\fN(\vr)\rar\l^{-1}\fN(\vr/\l)$, the
potential Lagrangian density $\L\_P(\vr)\rar\l^{-3}\L\_P(\vr/\l)$,
and so $\LP$ and the potential action are invariant. This means that
the field equations in the deep-MOND limit of the theory are
invariant to space scaling: If $\f(\vr)$ is a solution for
$\r(\vr)$, then $\f(\vr/\l)$ is a solution for $\l^{-3}\r(\vr/\l)$.
Recall that the deep-MOND limit of the nonlinear Poisson equation is
invariant to the whole group of conformal transformations in space,
of which scaling is one (Milgrom 1997).
\subsection{Virial relation}
If the system is everywhere deep in the MOND regime, for which
$\az^2\Q=(4/3)\n g^2\_N$ with our choice of the additive
normalization of $\Q$, the volume integral in eq.(\ref{virqet})
vanishes, and we have
 \beq \langle\vv^2\rangle={2\over 3}(MG\az)^{1/2}.
 \eeqno{verty}
This is an even more useful virial relation than eq.(\ref{virqet}),
since it relates the 3-D root-mean-square velocity, of any system of
test particles deep in the MOND regime, to the total mass,
irrespective of the mass distribution.
 \par
In many instances we want to view the system as made up of several
finite, non-test masses, ignoring the internal goings on in each of
them, and using only the center of mass velocities of these masses
(such as galaxies in a small group, or a binary). When the
accelerations of these masses are still deep in the MOND regime we
can write the above relation (following e.g., the argumentation in
Milgrom 1997) as
 \beq \langle\vv^2\rangle={2\over 3}(MG\az)^{1/2}(1-\sum_i
 q_i^{3/2}), \eeqno{virial}
where $q_i=m_i/M$. This is based on the relation
 \beq \sum_i \vr_i\cdot\vF_i=-{2\over
3}M(MG\az)^{1/2}(1-\sum_i
 q_i^{3/2}), \eeqno{virat}
 which applies in this case.
All these results are identical to those for the deep-MOND limit of
the nonlinear Poisson formulation. The general two-body force in the
deep-MOND limit follows straightforwardly from this relation, and is
thus the same as in the nonlinear Poisson formulation (see Milgrom
1997).
\par
Also discussed there is the relation between the above results and
the space scale invariance of the nonlinear Poisson formulation.
This is also relevant to the present theory. I only note here that
despite the formal invariance of $\LP$ under scaling, the fact that
$\LP$ itself is infinite leaves room for a finite change in $\LP$
(which can be viewed as the potential energy of the system) under
scaling of the density distribution $\r(\vr)\rar
\r_{\l}(\vr)=\l^{-3}\r(\vr/\l)$. So, under such a change in the mass
distribution we have
 \beq \LP[\r_{\l}]=\LP[\r]+ln(\l^{\V})=\LP[\r]+{2\over
3}M(MG\az)^{1/2}ln(\l). \eeqno{kirop}
 This violation of the
symmetry by the transformation properties of $\LP$ is analogous to
so called ``anomalies'' that appear in scale- or
conformally-invariant field theories, with the virial playing the
role of anomalous dimension. In a system made of finite, point
masses at positions $\vr_i$ we have\footnote{Here, we only move the
point masses to new positions $\l\vr_i$, without dilating them
intrinsically by a factor $\l$, as would be required for
eq.(\ref{kirop}) to apply.}
 \beq \LP(\l\vr_i)=\LP(\vr_i)+{2\over
3}M(MG\az)^{1/2}(1-\sum_i
 q_i^{3/2})ln(\l),  \eeqno{kirat}
giving rise to relation (\ref{virat}) by taking its $\l$
derivative at $\l=1$.

\section{A system in a constant external field}
\label{constant} Many times one is dealing with the dynamics in and
about a small system that is itself subject to the gravitational
field of a large,  mother system. Examples are the field around a
star, a globular cluster, or a dwarf companion in a field of a
mother galaxy. Instead of solving the full MOND
subsystem-plus-mother-system problem, we may want to treat the
subsystem alone, taking the presence of the mother system into
account through the constant-acceleration boundary condition that it
dictates in the vicinity of the subsystem. This approximation is
good if we can erect around the subsystem a volume $V$, whose
surface is $\S$, such that (i) $V$ is small enough that the
acceleration due to the mother system alone is nearly constant in
$V$, (ii) $V$ is large enough so that on $\S$ the perturbation due
to the subsystem is negligible compared with the acceleration due to
the mother system. Assume then that these conditions are satisfied.
Solve for the field of the mother system, which gives values
$\vgN_0$ and $\vg_0$ for the Newtonian and MOND accelerations at the
position of the subsystem. We do not (and need not) specify any
relation between these two accelerations. For a spherical mother
system they are related by the algebraic relation (\ref{algebraic}),
but in general they are not so related, and are not even parallel.
\par
Our approximation--becoming increasingly better as the mass of the
subsystem is small compared with that of the mother system--is to
solve the MOND field equations for the density $\r$ of the subsystem
alone, dictating as boundary conditions at infinity $\gf\rar-\vg_0$,
$\gfN\rar-\vgN_0$, instead of the conditions appropriate for an
isolated system. We write
 \beq \gfN= -\vgN_0+\grad\chi,
 ~~~~~~~~\gf= -\vg_0+\grad\psi,  \eeqno{huta}
and determine $\psi$ from eq.(\ref{poisa}) with the boundary
condition $\grad\psi\rar0$ at infinity, given that $\chi$ is the
standard Newtonian potential of the subsystem. Note, importantly,
that $\vg_0$ drops from the equation for the internal potential
$\psi$; so, this potential does not depend on $\vg_0$, only on
$\vgN_0$. Take the direction of the latter to define the positive
$z$ axis, and its value in units of $\az$: $\eN=|\vgN_0|/\az$.
\par
Unlike the isolated case, where $\hat M=\int\hr\drt$ is infinite,
for the case of a constant background field $\hat M$ is finite, and
is given by (Milgrom 2009a)
 \beq \hat M=-{1\over\fpg}\int\_{\infty}\n\vgN\cdot\vds=
 \n_0 (1+\hnz/3)M,  \eeqno{hutqa}
where $\n_0$ and $\hnz$ are, respectively, the values of $\n$ and
its logarithmic derivative at $\eN$  ($-1/2\le\hat\n_0\le 0)$, and
$M$ is the mass of the subsystem.
\subsection{Asymptotic behavior of the field}
The asymptotic behavior of $\hr$ is obtained from that of the
Newtonian field, and is given by
 \beq \hr\approx \n_0\hnz{M\over 4\pi}
 {1\over r^3}(1-{3z^2\over r^2}). \eeqno{muopl}
\par
The asymptotic internal MOND field $\psi$ satisfies the Poisson
equation for the above asymptotic form of $\hr$. This determines
$\psi$ up to a harmonic function (a solution of the Laplace
equation). Since $\psi$ is required to vanish at infinity, only
harmonic functions that decrease with $r$ have to be considered.
Among these we are interested in the leading term, which obviously
leaves us with a freedom to add to $\psi$ a multiple of $r^{-1}$.
The coefficient of this term is then determined by imposing
$\int\_{\S}\grad\psi\cdot\vds=\fpg \hat M$, which follows by
applying Gauss's theorem to eq.(\ref{poisa}). Because $\hr\propto
r^{-3}P_2(z/r)$ asymptotically, where $P_2$ is the second Legendre
polynomial, it follows that the sought after solution is of the form
$r^{-1}P_2(z/r)$. Matching coefficients, we get that
 \beq \psi^*\approx -{1\over 6}\n_0\hnz{MG\over
  r}(1-{3z^2\over r^2}) \eeqno{muoplui}
solves the above Poisson equation. As can be readily checked,
$\int\_{\S}\grad\psi^*\cdot\vds=0$ on a sphere. We thus have to add
to $\psi^*$ the potential $-\hat M G/r$ to satisfy Gauss's theorem.
This finally gives for the asymptotic form of $\psi$
 \beq \psi\approx -{MG\over r}\n_0(1+{\hnz\over 2})
 [1-{\hnz\over 2+\hnz}{z^2\over
 r^2}].  \eeqno{nuter}
\par
To compare our results with the analog ones for the nonlinear Poisson
formulation, which depend on the value of the MOND background
acceleration, we have to decide which value of $\vg_0$ to use when
comparing with the results for a given $\vgN_0$. I pick
heuristically, the value
 $\vg_0=\nu(|\vgN_0|/\az)\vgN_0$. We can then write
eq.(\ref{muopl}) as
 \beq \hr\approx -{M\over 4\pi\m_0}
 {L_0\over 1+L_0}{1\over r^3}(1-{3z^2\over r^2}), \eeqno{muoplaq}
where $\m_0$ and $L_0$ are, respectively, the values of $\m$, and of
its logarithmic derivative, at $\eta=|\vg_0|/\az$ [$\m(x)$ is
related to $\n(y)$ as described above]. For the same
 $\vg_0$, we can write eq.(\ref{hutqa}) as
 \beq \hat M={3+2L_0\over 3\m_0(1+L_0)}M,
   \eeqno{hutqasa}
 and the asymptotic potential as
  \beq \psi\approx -{MG\over \m_0 r}{1+L_0/2\over 1+L_0}
  \left(1+{L_0\over 2+L_0}{z^2\over
 r^2}\right).  \eeqno{nutola}
\par
These can now be compared with the analog quantities for the
nonlinear Poisson theory for a MOND external field $\vg_0$ (Milgrom
2009a):
 \beq \bar \r\approx  -{M\over 4\pi  \m_0}{L_0\over
 (1+L_0)^{3/2}}{1 \over \hat r^3}(1-{3\hat z^2\over \hat r^2}),
  \eeqno{mlyul}
where  $\hat z=(1+L_0)^{-1/2}z$, $\hat r=(x^2+y^2+\hat z^2)^{1/2}$.
We see a similar normalization and radial dependence as in $\hr$,
but a somewhat different angular dependence. For the total
``dynamical'' mass we have there
 \beq \bar M= {1\over \m_0 L_0^{1/2}}\arcsin\left({L_0\over
 1+L_0}\right)^{1/2}M,
\eeqno{asym} which, numerically, is not very different from
expression (\ref{hutqasa}), differing by at most 6 percent in the
possible range $0\le L_0\le 1$. For the asymptotic, internal
potential we have
 \beq \bar\psi\approx-{MG\over \m_0 r}(1+L_0)^{-1/2}
 \left(1-{L_0\over 1+L_0}{z^2\over
 r^2}\right)^{-1/2}.  \eeqno{yutred}
\par
An interesting difference between the two theories in the present
context is that the symmetry axis for the QUMOND theory is the
direction of the Newtonian background field, while that in the
nonlinear Poisson theory is the direction of the MOND background
field. The two directions are in general different. This shows that
the QUMOND theory cannot be equivalent to some form of the nonlinear
Poisson theory.\footnote{There is other evidence pointing to the
same effect: Consider, as an example, two unequal point masses. As
explained in Milgrom (1986b), a region where $\bar\r$ takes up both
signs appears around the critical point where $\gf=0$ between the
two masses. Using the same arguments, we now see that such a region
for $\hr$ appears where $\gfN=0$. The critical point for $\gf$ in a
nonlinear Poisson theory is never at the Newtonian critical point
(unless symmetry dictates it). So $\hr$ cannot be a $\bar\r$ for
some nonlinear Poisson theory. We shall see more arguments to this
effect below. If the present theory is equivalent to some nonlinear
Poisson formulation with some interpolating function $\m$, we would
have to have $\m(x)=1/\n(y)$, where $x\m(x)=y$ in order for the two
to coincide for spherical systems. But with this identification of
$\n$ the two theories are definitely not equivalent for aspherical
systems.} This difference can also help distinguish observationally
between the two types of theories.
\subsection{A system dominated everywhere by an external field}
\par
If  $|\grad\chi|\ll |\vgN_0|$ everywhere in the subsystem, we can
write $\vg=\vg_0-\grad\psi$, with $\vg_0$ dictated by the mother
system, and $\psi$ satisfying to lowest order in $\chi$
 \beq
 \Delta\psi=\n_0[\chi\_{,xx}+\chi\_{,yy}+(1+\hnz)\chi\_{,zz}]=
 \fpg\n_0\r+\n_0\hnz\chi\_{,zz},
 \eeqno{distar}
or, in terms of the above, heuristically chosen, $\vg_0$
  \beq
 \Delta\psi= \m_0^{-1}[\chi\_{,xx}+\chi\_{,yy}
 +(1+L_0)^{-1}\chi\_{,zz}].
 \eeqno{difer}
The analog equation for the internal potential in a dominant
external field, in the nonlinear Poisson formulation is (Milgrom
1986a)
 \beq[\psi\_{,xx}+\psi\_{,yy}+(1+L_0)\psi\_{,zz}]=
 {\fpg\over \m_0}\r=\m_0^{-1}\Delta\chi.
\eeqno{verut}
\par
In both theories the main effect is to enhance gravity by a factor
of $1/\m_0$ over Newtonian gravity. The secondary effect, that of
stretching the internal potential in the $z$ direction by a factor
of about $(1+L_0)^{1/2}$ is different in the two theories, as is the
direction of the $z$ axis itself.

\subsection{Center-of-mass acceleration of composite systems}
Consider a small body--such as an atom, a gas cloud, or a
star--freely falling in a mother system, such as a galaxy. Such
subsystems are made of constituents that are, sometimes, subject to
high internal accelerations, hence to high total accelerations. Is
it possible then that the center-of-mass motion of the composite
subsystem is subject to the MOND acceleration of the galaxy, as it
should for MOND phenomenology to work? This was shown to be the case
for the nonlinear Poisson formulation (Bekenstein \& Milgrom 1984),
and I now show that it is also the case for the present theory.
\par
Take an arbitrary, bounded mass distribution $\r$, representing the
subsystem, placed in a background MOND field $\vg_0$, such as a
galactic field in the example above.  As explained above, provided
the mass of the subsystem and its extent are much smaller than the
corresponding attributes of the galaxy, we can describe the MOND
field of the system by the solution of the MOND equations with a
boundary condition of constant acceleration at infinity.
\par
Use expression (\ref{juipo}) to calculate the force on the system as
an integral over the surface $\S$. Take $\S$ large enough so that on
it $\vg_0$ strongly dominates the internal acceleration, so we can
use the asymptotic form of the MOND field $\vg=\vg_0-\grad\psi$,
where $\psi$ is given by eq.(\ref{nuter}). Substituting this in the
integrand term $-\vg\vgN\cdot\vds$ in eq.(\ref{juipo}), the term
with $\vg_0$ gives $-\vg_0\int\_{\S}\vgN\cdot\vds=\fpg M\vg_0$,
where I used the Newtonian Gauss theorem. The term with $\grad\psi$
combined with all the other terms in eq.(\ref{juipo}) can be shown
to give a vanishing contribution in the limit where $\S$ goes to
infinity. We are then left with $\vF=M\vg_0$, independent of the
details of $\r$, giving a center-of-mass acceleration $\vg_0$.

\section{Discussion}
\label{discussion} I have presented a new NR formulation of MOND as
a modified gravitational potential theory. It is derivable from an
action and enjoys the standard conservation laws. In a sense, it is
an upgrade of the pristine formulation of MOND in which the MOND
acceleration field is an algebraic function of the Newtonian
acceleration field.
\par
The theory is a member in a class of bi-potential theories in which
only one--the MOND potential--couples to matter directly, whereas
the other is an auxiliary potential. The MOND departure from
Newtonian physics enters not through a modification of the free
action of the potential (as happens in the original formulation of
the Bekenstein Milgrom theory), but through the interaction between
the two potential fields. In Milgrom (2002b) I described a membrane
model for MOND in the spirit of the well known membrane model of
gravity. In this model MOND departure from Newtonian gravity is
introduced through modified `elasticity' of the membrane (compared
with that corresponding to newtonian gravity). In the same vein, the
present class of theories gravity can be attributed to the existence
of two membranes, one on which matter live and with which alone it
interacts directly, the other membrane interacting with the first.
MOND effects are then introduced not as modified elasticity of the
membranes--which is normal, albeit with the auxiliary membrane
having negative elasticity--but through the interaction between the
membranes.
\par
These theories are now added to the nonlinear Poisson formulation of
MOND propounded a quarter century ago by Bekenstein \& Milgrom
(1984), to which it is similar in many ways. The two theories are
shown, in fact, to fork out of the same MOND modification of a
Palatini-like formulation of Newtonian gravity.
\par
This addition carries with it the usual benefits of diversity. For
example, it brings home the possibility that even more formulations
await discovering, and encourages us to look for them.
\par
Also, comparing the predictions of different formulations helps
pinpoint results that are not generic to the MOND paradigm, not even
to its NR formulations. It may also point to new directions for
constructing relativistic formulations of MOND.
\par
In itself the new formulation is a complete theory, on a par with
the nonlinear Poisson formulation, but rather easier to work with,
involving, as it does, only linear differential equations.
\par
Because of the similarities between the two formulations, this
QUMOND formulation can also double as a well motivated approximation
for the less wieldy nonlinear Poisson formulation (and so also, for
the NR limit of theories of the class of TeVeS whose NR limit is a
double-potential theory of the nonlinear Poisson type). In fact, it
is exactly in this role that such a formulation was used in Milgrom
(2009a)--without recognizing its completeness as a theory--to
calculate the external-field effect in the inner solar system in the
nonlinear Poisson formulation. This yielded a closed expression for
the desired effect, whereas the application of the nonlinear theory
required solving the field equation first. Those ``approximate''
results are now recognized as exact in the new
theory.\footnote{There were also instances in the past, such as in
Milgrom (1986b), and in Milgrom \& Sanders (2008) where the present
formalism was used to approximate the `phantom' density of the
nonlinear Poisson theory; these too are exact in the present
theory.}

\section*{Acknowledgements}
This research was supported by a
center of excellence grant from the Israel Science Foundation.
\appendix
\section{Alternative expressions for the force the torque
and the virial}
\label{A}The force on a subsystem made of the masses within the
volume $V$ of surface $\S$, given in eq.(\ref{force}) can also be
written as
 \beq \vF=-{1\over\fpg}\int_V\drt\Delta\fN\gf. \eeqno{juote}
 The torque from eq.(\ref{opiona}) can be written as
 \beq \vT=-\int_V\drt\r\vr\times\gf=
 -{1\over\fpg}\int_V\drt\Delta\fN\vr\times\gf,\eeqno{juoteto}
and similarly for the virial. Integrating by parts several times in
a certain order, using the field equation (\ref{poisa}), and
expressing volume integrals of divergences as surface integrals, we
get expressions (\ref{juipo})(\ref{juipom})(\ref{moret}) for these
three quantities. Performing the integration by parts in a different
order we get other expressions for these quantities, which are also
useful
 \beq \vF=-{1\over \fpg}\int_{\S}\{\vg\vgN\cdot\vds
  +({\az^2\over 2}\Q-\n g\_N^2)\vds
 +\fN[(\vds\cdot\nabla)
 \vg-\grad(\n\vgN\cdot\vds)]    \}, \eeqno{muita}
or, writing, as in eq.(\ref{vure}), $\vg=\n\vgN+\curl\vA$ we get
 \beq \vF=-{1\over \fpg}\int_{\S}\{ \n\vgN\vgN\cdot\vds+
 ({\az^2\over 2}\Q-\n g\_N^2)\vds+(\curl\vA)\vgN\cdot\vds
 +\fN\grad[(\curl\vA)\cdot\vds]\}. \eeqno{muigy}
$$\vT={1\over \fpg}\int_{\S}\{-\vr\times\vg(\vgN\cdot\vds)
-{\az^2\over 2}\Q\vr\times\vds +\n\vr\times\vgN( \vgN\cdot\vds)+$$
 \beq +\fN[\div(\n\vgN)(\vr\times\vds)+\vg\times\vds
 -(\vr\times\nabla)(\vg\cdot\vds)]    \}. \eeqno{moya}

 $$ \fpg\V=\int\drt[2\n g^2\_N-{3\over
 2}\az^2\Q]+\int\vr\cdot\vg(\vgN\cdot\vds)+{1\over 2}\int[\az^2\Q-2\n
 g^2\_N]\vr\cdot\vds+$$
 \beq+\int\fN[(\vg-\n\vgN)\cdot\vds+(\vr\cdot\nabla)
 (\vg-\n\vgN)\cdot\vds].   \eeqno{nireq}
In all the above expressions $\grad$ in the last term does not act
on the surface element $\vds$.

\clearpage

\end{document}